\documentclass{emulateapj}
\usepackage{textcomp}
\usepackage{mathtext}
\usepackage{amssymb}

\newcommand{\fermi}{\textit{Fermi}}
\newcommand{\gr}{$\gamma$-ray}
\newcommand{\psr}{B1259$-$63}

\begin{document}

\title{\textit{Fermi} Study of 5--300 GeV emission from the high-mass pulsar binary PSR B1259$-$63/LS~2883}

\author{Yi Xing\altaffilmark{1}, Zhongxiang Wang\altaffilmark{1}, \& Jumpei Takata\altaffilmark{2}}

\altaffiltext{1}{Key Laboratory for Research in Galaxies and Cosmology, Shanghai Astronomical Observatory, Chinese Academy of Sciences,
80 Nandan Road, Shanghai 200030, China}
\altaffiltext{2}{Physics Department, Huazhong University of Science and Technology, Wuhan, China}

\begin{abstract}

We report the results from our detailed analysis of the \textit{Fermi} Large
Area Telescope (LAT) data for the pulsar binary PSR~B1259$-$63/LS~2883.
During the GeV flares that occurred when the pulsar was in the periastron passages,
we have detected a 5--300~GeV component at $\simeq 5\sigma$
in emission from the binary. The detection verifies the presence
of the component that has been marginally found in the previous studies of
the binary.  Furthermore, we have discovered that this component was present
even in the quiescent state of the binary, specifically the mean anomaly phase 0.7--0.9.
The component can be described by a power law with photon index $\Gamma\sim 1.4$,
and the flux in the flares is approximately one order of magnitude higher than that in quiescence.
We discuss the origin of this component. It likely arises from the inverse-Compton process:
high-energy particles from the shock, due to the interaction between the winds from the pulsar
and massive companion, scatter the seed photons from the companion to GeV/TeV energies.
Based on this scenario, model fits to the broad-band X-ray--to--TeV spectra of the binary
in the flaring and quiescent states are provided.

\end{abstract}


\section{Introduction}

Among a handful high-mass \gr\ binaries currently found in our 
Galaxy (e.g., \citealt{dub13,3fgl15}), PSR\ \psr/LS~2883 (hereafter \psr)
is unique in that 
its compact star is known being a $\simeq$47.8 ms radio pulsar \citep{joh+92},
while those in the others are not certain. Whether there is a black hole 
or a pulsar
in such a binary is critical for our understanding of their high-energy emission
mechanisms. For the former case, \gr\ binaries would be like microquasars 
with emission powered by accretion energy,
while for the latter, they would be pulsar binaries powered by 
the rotational energy \citep{dub13}.
Given that \psr\ might be the representing example of the \gr\ binary
group, providing a clear picture for the physical processes that occur
in such binaries, its general properties are of great importance and have
been studied in amazing detail at every accessible energy band.

The companion LS~2883 in \psr\ has been identified as an O9.5Ve main sequence 
(e.g., \citealt{joh+94,sjm14}) at a distance of 
$2.3\pm0.4$~kpc \citep{neg+11}. 
The pulsar moves around the massive companion with an orbital period of
$P\simeq 3.4$\,yr, and the orbit is highly eccentric ($e\sim 0.87$).
An orbital inclination angle of 24\fdg7 can be estimated from 
the mass function of this binary ($f_M = 1.53 M_{\odot}$, and assuming a 
standard neutron star mass of 1.4~$M_{\odot}$; \citealt{joh+94}).
It is considered that the pulsar wind has a strong 
interaction with the stellar wind, resulting in a termination shock, 
where particles are accelerated and nonthermal emission is radiated over a wide 
range of energy bands (e.g., \citealt{ta97}). 
As the pulsar moves near the periastron of the orbit, because
the equatorial disk of LS~2883 is inclined at an angle of 
$\mathrm{35^{o}}$ to the orbital plane \citep{mjm95,sjm14}, 
the pulsar crosses the disk twice.
During the crossing, much enhanced emission has been observed 
(e.g., \citealt{che+06,che+14}).  

Previously, \psr\ has been well studied at X-ray energies (see, e.g., 
\citealt{che+06,che+09}), and the very-high-energy (VHE) \gr\ emission
was discovered by High Energy Stereoscopic System (HESS; \citealt{aha+05}) 
during the 2004 periastron passage. After 
{\it Fermi Gamma-Ray Space Telescope (Fermi)} was launched in 2008,
GeV \gr\ emission was also detected during the 2010 periastron passage 
\citep{abd+2011}. Since then, more intensive studies
of the binary have been conducted over the past few years.
Now from various observations, it has been established that
during a periastron passage, the source's X-ray emission is 10 times 
enhanced and the flux
peak appears approximately 20 days after the periastron. At GeV energies, 
a flare that can be significantly detected by \fermi\ occurs
$\sim$30--80 days after the periastron \citep{tam+15,cal+15,che+15},
with emission described by a power law with 
photon index $\Gamma\simeq 3$ \citep{tam+15,cal+15}. The X-ray plus 
low-energy \gr\ emission during the same passage time periods
has been suggested to be due to synchrotron radiation of high energy 
particles accelerated at the termination shock of the pulsar wind 
(e.g., \citealt{tt09,kon+11,che+15}). 
In addition, marginal detections of the source in the \fermi\ high 
energy range (several tens of GeV) during the flares were recorded 
(see \citealt{tam+15} and Figure 8 in \citealt{che+15}).
This component may be the low energy tail of the spectral component 
observed at TeV by HESS \citep{hess13}, and has been suggested to be due to 
the inverse Compton (IC) process: the stellar photons are up-scattered by
the particles from the shocked pulsar wind to the high energies. 
(e.g., \citealt{tt09,kon+11,che+15}).
\begin{figure}
\centering
\epsscale{1.0}
\plotone{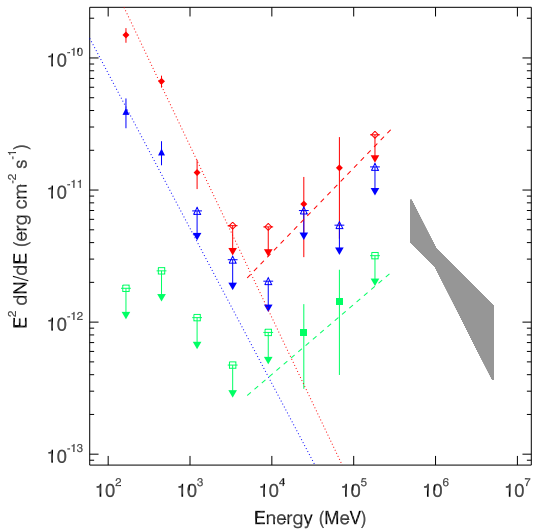}
\caption{\fermi\ \gr\ spectra of \psr\ during the flaring state (red diamonds),
periastron passage excluding the flares (blue triangles), and 
quiescent state (green squares). The red and blue dotted lines are 
the 0.2--300 GeV power-law fits during the flaring state and 
non-flaring state in the periastron passage, respectively.  The red and 
green dashed lines are the 5--300 GeV power-law fits during the flaring state 
and quiescent state, respectively. The grey area marks 
the power-low spectrum of \psr\ obtained with HESS \citep{hess13}.}
\label{fig:spectra}
\end{figure}

In this paper we report our detailed analysis of the \fermi\ Large Area 
Telescope (LAT) data of the binary for studying the high energy component.
From the analysis, we verified the detection of the component during 
the flares, and found that this component existed even during the quiescent 
time period (i.e., excluding the periastron passages).

\section{Data Analysis and Results} 
\label{sec:ana}

\subsection{\textit{Fermi} LAT Data}

LAT is a $\gamma$-ray imaging instrument onboard \fermi.
It scans the whole sky every three hours and can continuously 
conduct long-term \gr\ observations of thousands of GeV 
sources \citep{atw+09}. In the analysis, we selected 0.1--300 GeV LAT events 
from the \textit{Fermi} Pass 8 database inside 
a $\mathrm{20^{o}\times20^{o}}$ region centered at the position of \psr, 
which is R.A.=13$^{\rm h}$02$^{\rm m}$47$\fs$64, 
Decl.=$-$63$^{\circ}$50$'$08$\farcs$67 (equinox J2000.0; \citealt{sjm14}). 
The time period of the LAT data is 
from 2008-08-04 15:43:36 (UTC) to 2015-11-26 00:08:16 (UTC). Following 
the recommendations of the LAT 
team\footnote{\footnotesize http://fermi.gsfc.nasa.gov/ssc/data/analysis/scitools/}, 
we included the events with zenith angles less than 90 degrees, which 
prevents the Earth's limb contamination, and excluded the events with 
quality flags of `bad'.

\subsection{Likelihood and Spectral Analysis of the Periastron Passage Data}
\label{subsec:la}

We first repeated the analysis reported in the previous studies
of the periastron passages in year 2010 (MJD 55545) 
and 2014 (MJD 56781; \citealt{abd+2011,tam+11,tam+15,cal+15}).
Around the periastron time $T_0$, the passage period was considered to be 
from $T_0-50$ to $T_0+100$ days, within which the flares (when
the GeV emission was detected in a one-day timescale)
occurred in a time period $T_f$ of 30--80 days after $T_0$
(see, e.g., \citealt{tam+15,che+15,cal+15}). Excluding $T_f$ within
each of the periastron passages, 
the remaining non-flaring time period was defined as $T_n$.

\subsubsection{Likelihood Analysis}

We included all sources within 20 degrees centered at the position 
of \psr\ in the \textit{Fermi} LAT 4-year catalog \citep{3fgl15} to make 
the source model. The spectral forms of these sources are provided in 
the catalog. The spectral parameters of the sources within 5 degrees 
from \psr\ were set as free parameters, and the other parameters were 
fixed at their catalog values. \psr\ was included in the source model 
as a point source with emission modeled with a power law. In addition, 
we considered the background Galactic and extragalactic diffuse emission 
in the source model with the spectral model gll\_iem\_v06.fits and the 
file iso\_P8R2\_SOURCE\_V6\_v06.txt, respectively. The normalizations of 
the diffuse components were set as free parameters.
\begin{figure*}
\centering
\epsscale{1.0}
\plotone{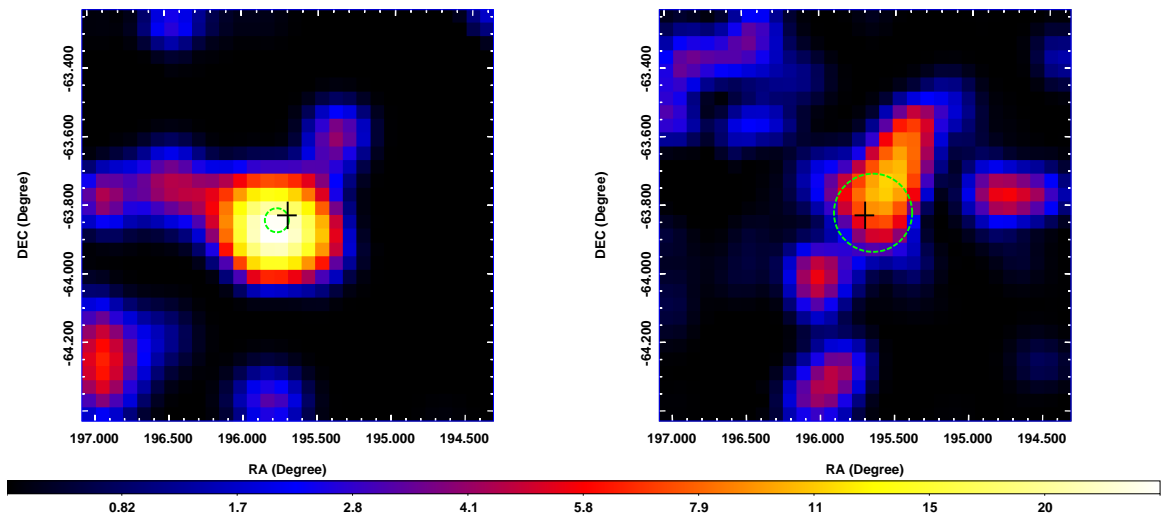}
\caption{TS maps of a $\mathrm{1.2^{o}\times1.2^{o}}$ region
centered at \psr\ in the 5--300 GeV band, during the flaring state 
({\it left} panel) and quiescence ({\it right} panel). The image scale 
of the maps is 0\fdg04 pixel$^{-1}$. The color bar indicates the TS value 
range. All catalog sources were considered and removed, and no catalog sources
are within the square region. The dark crosses mark the position of \psr. 
The green circles mark the 2$\sigma$ error circles of the best-fit positions 
obtained in each state.}
\label{fig:tsmap-state}
\end{figure*}

Using the LAT science tools software package {\tt v10r0p5},
we performed standard binned likelihood analysis to the LAT data in 
the $>$0.2 GeV band.
The events below 200 MeV were not included, so as to reduce the effects of 
the relatively large uncertainties of the instrument response function of 
the LAT in the low energy range. 
We found photon index $\Gamma= 3.3\pm$0.1 and photon 
flux $F_{0.2-300}= 1.26\pm 0.09\times 10^{-7}$ photons~s$^{-1}$\,cm$^{-2}$, 
with a Test Statistic (TS) value of $\sim$406 for \psr\ in the periastron 
passages.  A TS value at a given position 
is calculated from TS$= -2\log(L_{0}/L_{1})$, where $L_{0}$ and $L_{1}$ 
are the maximum likelihood values for a model without and with an additional 
source respectively. It is a measurement of the fit improvement for including 
the source, and is approximately the square of the detection significance 
of the source \citep{1fgl}. 

In addition, we also performed likelihood analysis during $T_f$ and $T_n$.
We found $\Gamma= 3.34\pm$0.04, 
$F_{0.2-300}= 2.5\pm 0.1\times 10^{-7}$ photons~s$^{-1}$\,cm$^{-2}$ 
(with a TS value of $\sim$545) in the former, and
$\Gamma= 3.2\pm$0.3,
$F_{0.2-300}= 5.0\pm 1.0\times 10^{-8}$ photons~s$^{-1}$\,cm$^{-2}$ 
(with a TS value of $\sim$42) in the latter.
The results in $T_f$ are consistent with those in the previous studies 
(see \citealt{cal+15,che+15}). The $T_f$ flux is
5 times higher than that 
during $T_n$, while the photon indices in the two time periods are 
approximately the same.

\subsubsection{Spectral Analysis}
\label{subsec:sa}

We extracted the $\gamma$-ray spectra of \psr\ during 
$T_f$ and $T_n$
by performing maximum likelihood analysis of the LAT data in 8 evenly 
divided energy bands in logarithm from 0.1--300 GeV. 
In the extraction, the spectral normalizations of the sources within 5 
degrees from \psr\ were set as free parameters, while all the other parameters 
of the sources were fixed at the values obtained from the above maximum 
likelihood analysis. We kept only spectral data points when TS greater 
than 4 ($>$2$\sigma$ significance) and derived 95\% flux upper limits 
otherwise. 

The obtained spectra are shown in Figure~\ref{fig:spectra}, with
the flux and TS values of the spectral data points provided 
in Table~\ref{tab:spectra}. 
In the spectra, only emission at 0.16 and 0.45 GeV bands 
in $T_n$ was significantly detected, and a high-energy component 
was detected in $T_f$ as the TS values at 25 and 67 GeV bands are 12 and 14 
respectively (see Table~\ref{tab:spectra}).

\subsection{Analysis for the High-Energy Component}

\subsubsection{Likelihood Analysis}

To study the properties of the high-energy component, 
we first performed binned likelihood analysis of the LAT data 
in the $>$5 GeV energy
range. We found TS$\simeq$26 in $T_f$, where $\Gamma= 1.4\pm$0.4 
and 5--300\,GeV photon flux 
$F_{5-300}= 6\pm 3\times 10^{-10}$ photons~s$^{-1}$\,cm$^{-2}$
were obtained.
We also performed the analysis of the data in the `quiescent' time 
period $T_q$, i.e., excluding the periastron passages of $T_0-50$ 
to $T_0+100$ days, and
found TS$\simeq$11, where $\Gamma= 1.5\pm$0.4 and 
$F_{5-300}= 6\pm 3\times 10^{-11}$ photons~s$^{-1}$\,cm$^{-2}$ were
obtained.
The results indicate $\sim 5\sigma$ and $\sim 3\sigma$ detections of
the component in the two time periods. The component probably 
has the same power-law emission all the time, while the $T_f$ flux was one 
order of magnitude higher than 
that in $T_q$. We tested the analysis of the data in $>$10 GeV or
$>$15 GeV energy ranges. However the results were not as significant as that
in $>$5 GeV energy range.

The $>$5 GeV TS maps of a $\mathrm{1.2^{o}\times1.2^{o}}$ region 
centered at \psr\ during the two time periods are shown in
Figure~\ref{fig:tsmap-state}. 
No catalog sources are within the square region. The \gr\ emission 
appears to be a typical point source
in the $T_f$ TS map, but somehow slightly extended towards the north-west
direction in $T_q$.

Running \textit{gtfindsrc} in the LAT software package, we determined 
their positions. For the flaring state, the best-fit position 
is R.A.=195\fdg77, Decl.=$-$63\fdg85 (equinox J2000.0), 
with 1$\sigma$ nominal uncertainty of 0\fdg02. \psr\ is 0\fdg03 away from 
the position, outside of the 1$\sigma$ but within the 2$\sigma$ 
error circles.  For the quiescent state, R.A.=195\fdg65, Decl.=$-$63\fdg83 
(equinox J2000.0), with 1$\sigma$ nominal uncertainty of 0\fdg07. 
\psr\ is 0\fdg02 away from this best-fit position and within 
the 1$\sigma$ error circle.

\subsubsection{Orbital Variability}
\label{subsec:ov}

\begin{figure}
\centering
\epsscale{1.0}
\plotone{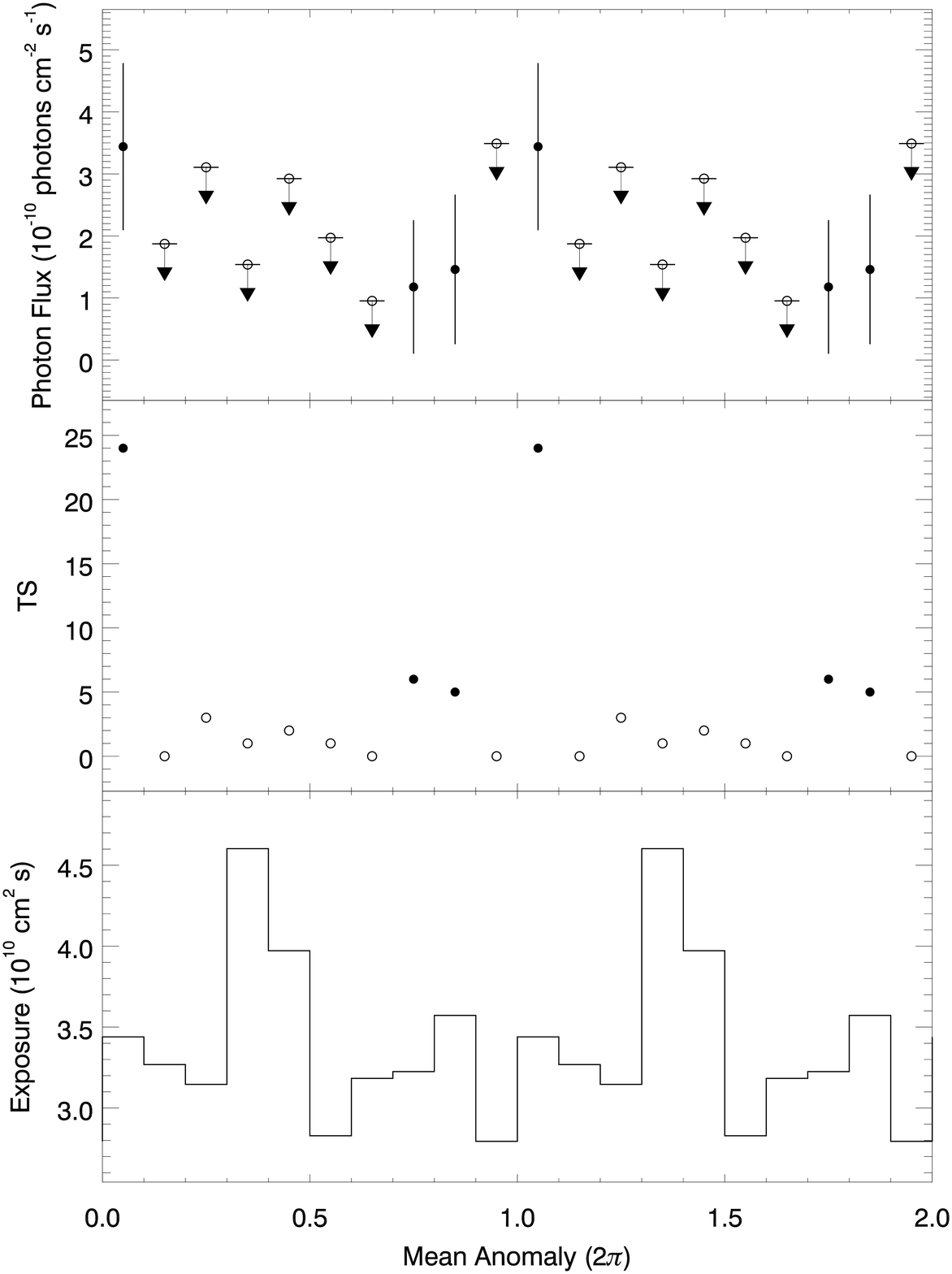}
\caption{Orbital (mean anomaly) curves of the 5--300 GeV photon flux 
({\it top} panel) and TS value ({\it middle} panel). The exposure times for
each phase bin are shown in the {\it bottom} panel.
Two cycles are displayed for clarity. }
\label{fig:profile}
\end{figure}

To further constrain the presence of the high-energy component in $T_q$, 
we performed likelihood analysis of the $>$5 GeV data in 10 orbital phase bins.
Because of the highly eccentric orbit of \psr, we used the mean anomaly phase 
bins so as to have approximately similar exposures in every bin. 
We set the spectral normalizations of the sources within 5 degrees 
from \psr\ as free parameters, while all the other parameters of the sources 
were fixed at the values obtained from the likelihood analysis of 
the $>$5 GeV data. The orbital light curve and TS curve extracted are shown in 
Figure~\ref{fig:profile}.  The phase zero was set at $T_0$. 
Besides the data point detected in phase 0.0--0.1 that is over part of
the periastron passage (each phase bin corresponds to $\sim$120 days), 
two data points have TS values greater than 4 in phase 0.7--0.8 and 0.8--0.9. 

The binary has a long orbital period, and therefore we checked 
the exposure time for each bin. We binned $>$5 GeV events 
in a 0\fdg3 radius circular region
(the size of the LAT's Point Spread Function at 5 GeV) 
centered at \psr\ in 1-day time bins, and calculated the exposures of 
the time bins using \textit{gtexposure} in the LAT software package. 
The obtained exposure time curve is plotted in Figure~\ref{fig:profile}
as a reference. It shows that the exposures have a maximum factor of 1.6 
differences. Our analysis results
are not affected by the exposure differences, since the phase ranges of 
0.7--0.9 do not have the highest exposure time.

\begin{figure*}
\centering
\epsscale{1.0}
\plotone{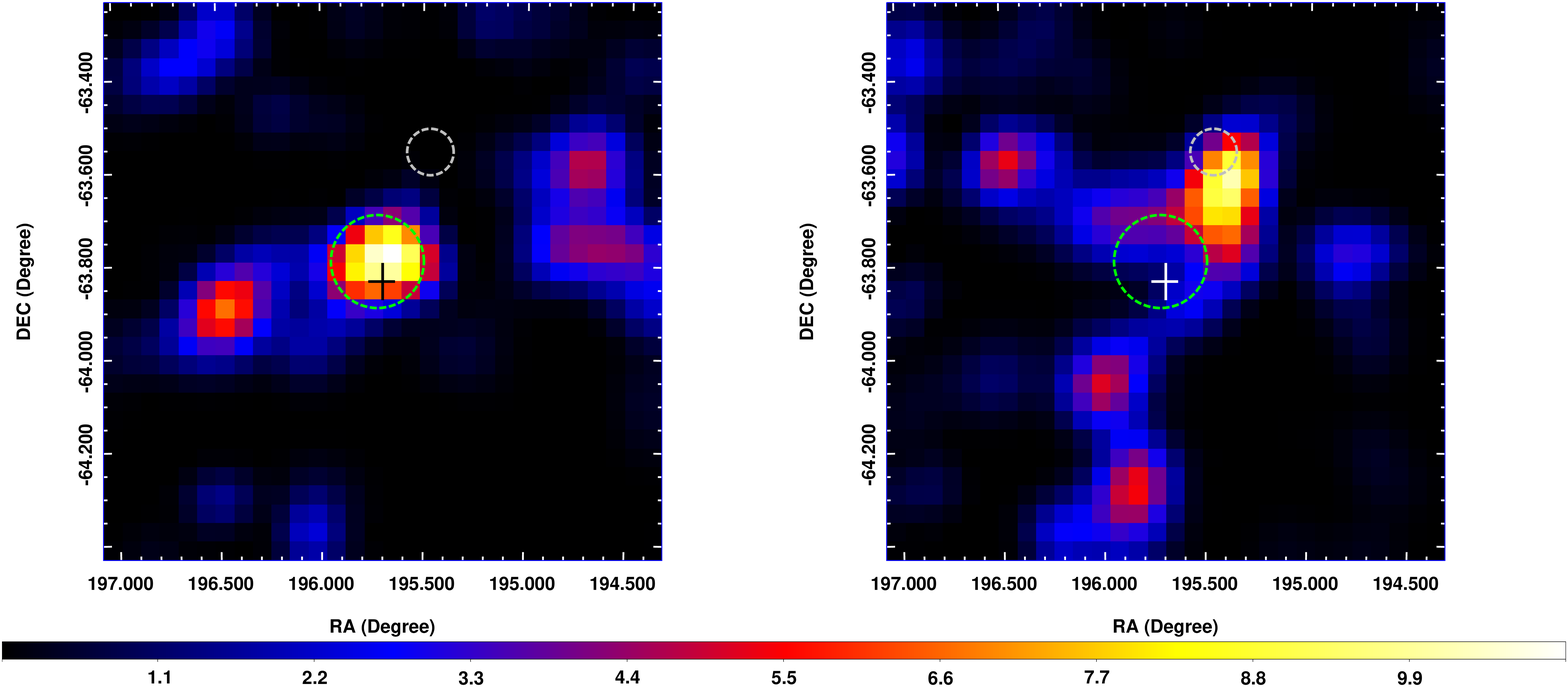}
\caption{TS maps of a $\mathrm{1.2^{o}\times1.2^{o}}$ region centered 
at \psr\ in the 5--300 GeV band, during phase 0.7--0.9 ({\it left} panel) 
and 0.1--0.7 ({\it right} panel). The image scale of the maps 
is 0\fdg04 pixel$^{-1}$. The color bar indicates the TS value range. 
The dark or white crosses mark the position of \psr. The green and gray 
circles mark the 2$\sigma$ error circles of the best-fit positions for
the sources detected 
during phase 0.7--0.9 and 0.1--0.7, respectively.}
\label{fig:tsmap-phase}
\end{figure*}

We calculated two TS maps during phase 0.7--0.9 and 0.1--0.7 to confirm
the light curve result, which
are shown in Figure~\ref{fig:tsmap-phase}. The high-energy emission during 
phase 0.7--0.9 was detected with TS of $\sim$11. 
The best-fit position 
is R.A.=195\fdg72, Decl.=$-$63\fdg79 (equinox J2000.0), 
with 1$\sigma$ nominal uncertainty of 0\fdg06. \psr\ is 0\fdg05 from 
the best-fit position and within the 1$\sigma$ error circle. 
During phase 0.1--0.7, TS$\sim$2, indicating the non-detection of the component.
However in the latter TS map, a nearby additional source was 
possibly revealed at the north-west of \psr\ with TS$\simeq$10. 
We determined the best-fit position 
for the additional source and obtained R.A.=195\fdg47, Decl.=$-$63\fdg56 
(equinox J2000.0), with 1$\sigma$ nominal uncertainty of 0\fdg03.

\subsubsection{Spectral Analysis}

We extracted the \gr\ spectrum in $T_q$, which is shown in 
Figure~\ref{fig:spectra}.  There was no detection of any 
low-energy emission ($\lesssim 9$\,GeV), consistent with the results 
from previous studies. The high-energy emission was marginally detected at 
25 and 67 GeV bands (the TS values are 5; Table~\ref{tab:spectra}).  

The origin of the nearby source found in phase 0.1--0.7 is not clear, 
and since it is located close to the \psr\ counterpart 
(outside of the 2$\sigma$ error 
circle; see the right panel of Figure~\ref{fig:tsmap-state}) 
determined in phase 0.7--0.9, there could be possible contamination due to
their proximity. To check this, we performed likelihood analysis to 
the $>$5 GeV LAT data during phase 0.7--0.9 when the nearby source
is not significantly present, and extracted the spectrum during the phase 
range. The results are consistent with those obtained for the whole 
quiescent time period.

\section{Discussion}
\label{sec:disc}

Having analyzed more than 7 years of \fermi\ LAT Pass 8 data for 
the \psr\ region, we have verified that in addition to the low-energy 
\gr\ emission with $\Gamma \sim$3, a high-energy component
in $>$5 GeV energy range was clearly present during the flaring state. 
This component was only marginally detected ($< 3\sigma$) in the previous 
studies of
the pulsar's periastron passages (e.g., \citealt{tam+15,che+15}).
The improvement on the detection significance is likely due to the use
of Pass 8 data in our analysis. 
Moreover, our detailed analysis indicates that this component even existed 
during the quiescent state, mean anomaly phase 0.7--0.9. The brightening 
of the component during the flaring state, 
with the flux an order of magnitude larger than that during quiescence, 
helps confirm the weak detection ($\simeq 3\sigma$) in the latter state. 
In addition,
a nearby source, north-west to \psr, was likely revealed from our analysis.
The existence of this source is probably the reason why \psr\ appeared 
slightly extended along the north-west direction in the quiescent 5--300~GeV
TS map (right panel of Figure~\ref{fig:tsmap-state}).

The $>$5 GeV emission component during quiescence is only significant
in the mean anomaly phase of 0.7--0.9, corresponding to $-$371 
to $-$123 days before the periastron. Multiple X-ray observations have
shown that for the phase range before the pulsar's entrance to 
the circumstellar disk, fluxes were low with $\Gamma\sim 1.8$ \citep{che+06}.
At TeV energies, the earliest HESS detection reported probably was
from $-$104 days from the periastron (2.9$\sigma$ detection from a 5.3 hour
livetime observation; \citealt{aha+09}). Towards the periastron, the 
TeV flux gradually increased and then rose to a level of more than 10 times 
larger during the periastron passage \citep{aha+09,che+14}.
We note that our detection of $>$5 GeV emission during phase 0.7--0.9 is 
only 3$\sigma$, and the \fermi\ exposure for phase 0.9--1.0 is the lowest 
(see Figure~\ref{fig:profile}). Thus we suspect that the $>$5 GeV emission,
as the low energy tail of the TeV emission, would start increasing 
from phase 0.7, and might become detectable in phase 0.9--1.0 in the near 
future when sufficient \fermi\ data are collected.
Since the mean anomaly phase 0.7 corresponds to the true anomaly phase 
0.53, the high energy component is present in most of the second half 
of the binary orbit.
\begin{figure}
\centering
\epsscale{1.0}
\plotone{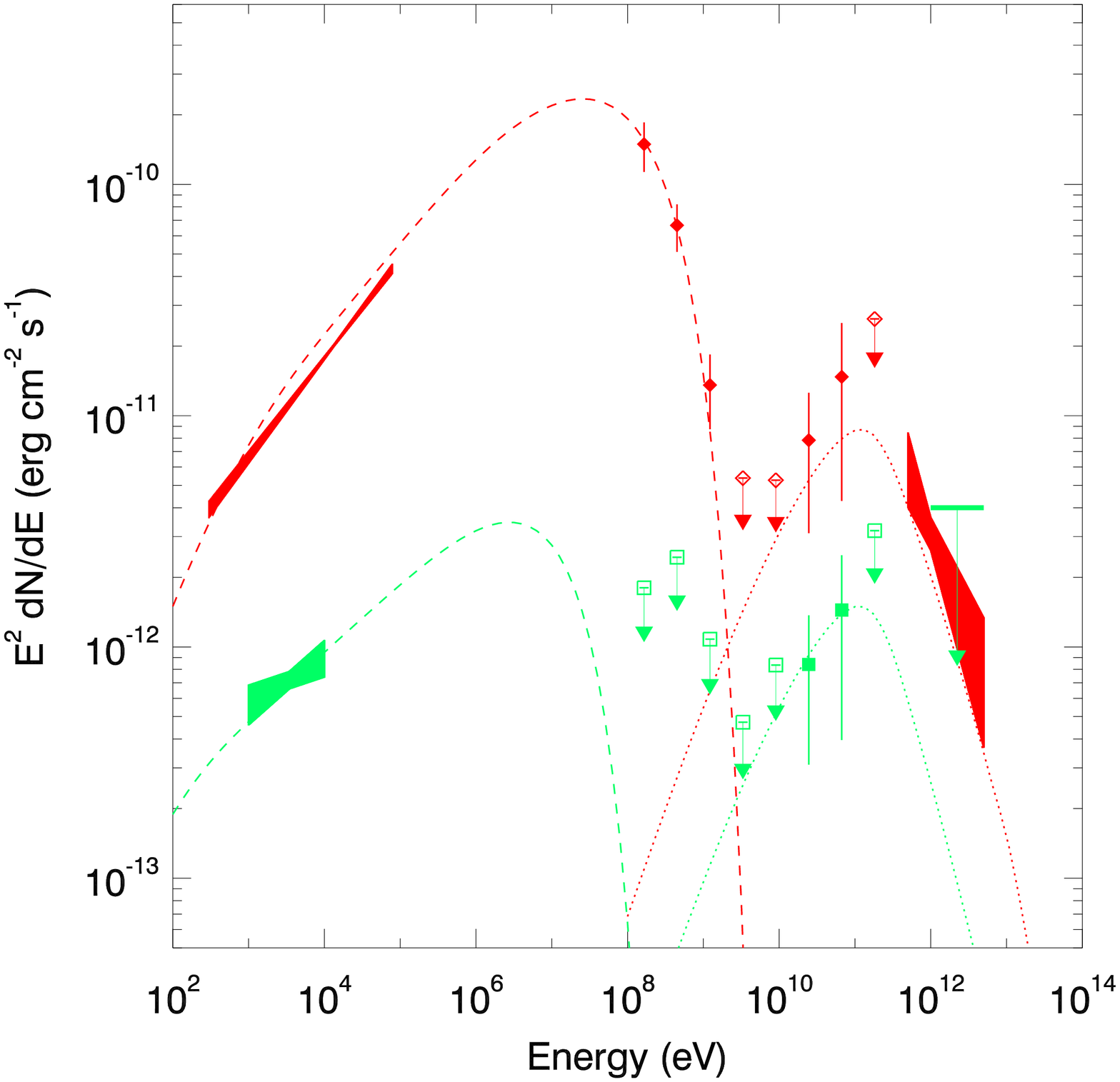}
\caption{Broad-band spectra of \psr\ in the flaring (red) and quiescent
(green) states and our model fits to them. The X-ray spectra in the former 
and latter state
are the \textit{NuSTAR}+\textit{Swift} data (2014 June 2; \citealt{tam+15}) 
and \textit{Swift} data (2010 August, $-127$ days before $T_0$;
\citealt{che+14}).
The TeV spectrum is from the 2011 periastron
passage \citep{hess13}, and the upper limit on the TeV flux in quiescence
is estimated by assuming the value to be lower than the 2007 May measurement
(see Table 2 \& 3 in \citealt{aha+09}).  
}
\label{fig:sed}
\end{figure}

With the low-energy GeV spectrum of \psr\ having $\Gamma \sim 3$, it is likely
that the  $<1$~GeV emission is the high-energy tail of the synchrotron
emission from the shocked pulsar wind, which explains the observed X-ray
emission. On the other hand, the observed emission
above 5~GeV is the low-energy tail of the IC process, which
produces TeV emission observed at the periastron passage, of the shocked
pulsar wind off the stellar photons. Based on this scenario,
we fit the broadband X-rays--to--TeV spectra with the model that includes
the synchrotron radiation
and IC scattering process of the shocked pulsar wind.
The model is based on \citet{tt09}. We assumed that the energy
distribution of the shocked particles forms a broken power-law function at
the shock, and the Lorentz factor at the break in the distribution is
$\Gamma_{b}\sim 10^6$, below and above which we fixed the power law
indices $p_1$ and $p_2$ at $p_1=1$ and $p_2=2.3$ (or $p_2=2.4$ when in
$T_q$), respectively. 
The maximum Lorentz factor of the particles accelerated at the shock was assumed
to be that from balancing between
the synchrotron radiation time scale $\tau_{syn}=9m_e^3c^5/4e^4B^2\Gamma$
and the acceleration time scale $\tau_{acc}=\xi \Gamma m_ec/eB$ 
(where $\xi\sim 1$; see \citealt{tt09} for the details).
We found that the efficient acceleration process
$\xi\sim 0.5$ was needed to explain the \fermi\ observation below 1~GeV with
the synchrotron radiation process. We calculated the evolution of
the particle distribution with the adiabatic loss and radiation losses of
the synchrotron radiation and the IC scattering process.
The orbit of \psr\ is highly eccentric. Hence, we used
the typical value of the separations between the pulsar and companion
to fit the observed phase-averaged spectra in the two states; $R_p=7$~AU
for the quiescent state and $R_p=2$~AU for the flaring state.
The shock distance from the pulsar was calculated from the wind momentum
ratio $\eta=0.2$, which is the typical value of the momentum ratio of
the pulsar wind and stellar wind/disk. The model fits to the spectra 
are shown in Figure~\ref{fig:sed}. The spectra can generally be described
by the model.

Since the detection of the orbital variability in X-ray 
(e.g., \citealt{che+06,che+09}) and TeV \citep{aha+09} fluxes, detailed,
different models \citep{kha+07,tt09,kon+12} have been proposed to explain the
observed orbital properties.  Our detection of the 5--300~GeV emission
in the quiescent phase 0.7--0.9 has provided another piece of observational
result to be
considered in modeling. As the IC process should occur over the whole orbit,
the high-energy emission is always being radiated. However in phase 0.1--0.5
when the pulsar moves towards us, the IC scattering between particles 
that moves toward us and stellar photons is a tail-on collision process, 
which significantly reduces the radiation efficiency. We suspect that
this difference would probably be the
reason why we do not seen the component in the first half of the orbit. 
In order to fully explore this possibility, detailed modeling of 
the processes over the orbit is warranted.

\acknowledgements
This research made use of the High Performance Computing Resource in the Core
Facility for Advanced Research Computing at Shanghai Astronomical Observatory.
This research was supported by the Shanghai Natural Science 
Foundation for Youth (13ZR1464400), the National Natural Science Foundation
of China for Youth (11403075), the National Natural Science Foundation
of China (11373055), and the Strategic Priority Research Program
``The Emergence of Cosmological Structures" of the Chinese Academy
of Sciences (Grant No. XDB09000000). J.T. acknowledges the support 
by the NSFC grant of China under 11573010.

\begin{table}
\tabletypesize{\footnotesize}
\tablecolumns{10}
\tablewidth{240pt}
\setlength{\tabcolsep}{2pt}
\caption{\fermi\ LAT flux measurements of \psr}
\label{tab:spectra}
\begin{tabular}{lccccccc}
\hline
\multicolumn{2}{c}{ } &
\multicolumn{2}{c}{Flaring state} &
\multicolumn{2}{c}{Periastron passage} &
\multicolumn{2}{c}{Quiescent state}  \\
\multicolumn{2}{c}{ } &
\multicolumn{2}{c}{ } &
\multicolumn{2}{c}{excluding the flares} &
\multicolumn{2}{c}{}  \\ \hline
$E$ & Band & $E^2dN(E)/dE$ & TS & $E^2dN(E)/dE$ & TS & $E^2dN(E)/dE$ & TS \\
(GeV) & (GeV) & (10$^{-12}$ erg cm$^{-2}$ s$^{-1}$) &  & (10$^{-12}$ erg cm$^{-2}$ s$^{-1}$) &  & (10$^{-12}$ erg cm$^{-2}$ s$^{-1}$) &  \\ \hline
0.16 & 0.1--0.3 & 149.3$\pm$18.9 & 637 & 39.4$\pm$10.0 & 93 & 1.8 & 0 \\
0.45 & 0.3--0.7 & 66.6$\pm$6.9 & 267 & 19.4$\pm$4.0 & 44 & 2.4 & 1 \\
1.22 & 0.7--2.0 & 13.6$\pm$3.4 & 25 & 7.0 & 3 & 1.1 & 1 \\
3.32 & 2.0--5.5 & 5.4 & 1 & 3.0 & 0 & 0.5 & 0 \\
9.03 & 5.5--14.9 & 5.3 & 1 & 2.0 & 0 & 0.8 & 0 \\
24.58 & 14.9--40.5 & 7.8$\pm$4.7 & 12 & 7.0 & 2 & 0.8$\pm$0.5 & 5 \\
66.86 & 40.5--110.3 & 14.7$\pm$10.4 & 14 & 5.4 & 0 & 1.4$\pm$1.0 & 5 \\
181.89 & 110.3--300.0 & 26.2 & 0 & 15.0 & 0 & 3.2 & 0 \\
\hline
\end{tabular}
\vskip 1mm
\footnotesize{Note: fluxes without uncertainties are 
the 95$\%$ upper limits.}
\end{table}

\end{document}